# Influence of Γ-X mixing on optical orientation and alignment of excitons in (In,Al)As/AlAs quantum dots

S. V. Nekrasov[1,*], N. O. Mikhailenko[1], M. D. Ragoza[1], T. S. Shamirzaev[2] and Yu. G. Kusrayev[1]

[1]*Ioffe Institute, Russian Academy of Sciences, 194021 St. Petersburg, Russia*
[2]*Rzhanov Institute of Semiconductor Physics,*
*Siberian Branch of the Russian Academy of Sciences,*
*630090 Novosibirsk, Russia*

**Abstract.** *The effect of Γ-X mixing on the energy levels fine structure of indirect in k-space excitons in an ensemble of (In,Al)As/AlAs quantum dots with type I band alignment was experimentally studied. Using the methods of optical spin orientation and optical alignment in a magnetic field, an increase in the anisotropic exchange splitting of excitonic levels (from approximately 0.6 to 5 μeV) due to the Γ-X mixing was revealed. The extent of direct electronic states admixing to indirect ones depends on the size of the quantum dot. On the other hand, the optical and spin properties of excitons change radically with increasing of the Γ states admixture to the X states: in the absence of a magnetic field, the optical orientation of excitons decreases from 18 to 3%, while the alignment of excitons is restored from 6 to 53%.*

## 1. INTRODUCTION

The prospect of using semiconductor quantum dots (QDs) in quantum computing is promising; in particular, the spin of a carrier localized in a QD can be considered as an implementation of an information qubit [1, 2, 3]. However, from an applied point of view, the lifetime of a photoexcited exciton in direct band gap QDs is extremely short - on the order of a nanosecond [4]. The spin relaxation time turns out to be even shorter; usually, in epitaxial QDs, the anisotropic component of the exchange interaction of an electron and a hole erases the spin orientation of a neutral exciton in a time on the order of 10 ps [5, 6]. The use of indirect band gap QDs helps to solve the problem of the short exciton spin lifetime. In this work, indirect band gap in k-space semiconductor (In,Al)As/AlAs QDs are studied. They are interesting for the long lifetime of photoexcited excitons, which can reach milliseconds [7]. Moreover, in Ref. [8] it was experimentally demonstrated, that the exciton spin relaxation time exceeds its lifetime due to the insignificance of the anisotropic exchange splitting of an indirect exciton ($\delta_b \leq 0.2$ μeV).

(In,Al)As/AlAs QDs can have either direct or indirect band gap in momentum space depending on the size of the quantum dot [9], the energy minimum in the conduction band can be located in the Γ or X valley of the Brillouin zone, respectively. There is a size of the QDs at which electron energies at the Γ and X points coincide; we will call the exciton energy corresponding to this case the point of Γ-X crossing ($E_{\Gamma X}$, see Fig. 1). In a certain energy range near the crossing point, the phenomenon of Γ-X mixing of direct and indirect exciton states takes place due to the quasiparticle reflection from the QDs interfaces. In (In,Al)As/AlAs QDs, the presence of the Γ-X mixing made it possible to study the properties of an indirect in k-space exciton using resonant spin-flip Raman scattering [10], in particular, the g factor of the electron in the X valley was determined.

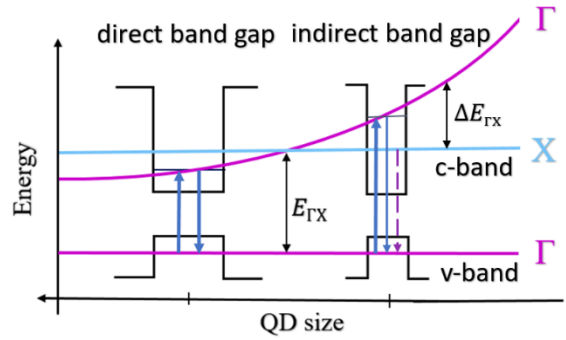

Fig. 1. Scheme of the band structure of (In,Al)As/AlAs QDs depending on the size of QDs. The Γ-X crossing energy ($E_{\Gamma X}$) is indicated.

In Ref. [11], the value of the direct excitons anisotropic exchange splitting $\delta_b$ depending on the size of the QD was measured using polarized photoluminescence (PL) spectroscopy of single (In,Al)As/AlAs QDs; for different QDs in the ensemble, $\delta_b$ was found in the range from 80 to 500 μeV.

In this work, we study the effect of the Γ-X mixing on the fine structure of nominally indirect excitons and, as a consequence, on the optical orientation and alignment of excitons in (In,Al)As/AlAs QDs. In excitons "purely" indirect in momentum space in (In,Al)As/AlAs QDs, the anisotropic exchange splitting does not exceed 0.2 μeV [8], thus it cannot be measured directly by the single QD PL spectroscopy used in Ref. [11]. In this work, we study the fine structure of indirect excitons using methods of optical orientation and alignment of excitons in a magnetic field in Faraday geometry. The measurements were carried out at different spectral points of the PL band, which correspond to different levels of mixing of direct exciton states with indirect ones.

## 2. EXPERIMENTAL DETAILS

The sample was grown by molecular beam epitaxy on a (001) GaAs substrate; the structure contains 20 layers of nominally undoped (In,Al)As/AlAs QDs. The average diameter and height of QDs are 15 nm and 4 nm, respectively. The density of quantum dots in each layer is approximately $3 \times 10^{10}$ cm$^{-2}$. The layers are separated from each other by 20-nm-thick AlAs barriers; the sample is protected from oxidation by a 20-nm-thick GaAs layer. More detailed characterization of similar samples can be found in Ref. [9].

An experimental setup for studying polarized PL in a magnetic field was used. The PL was excited be Ti:sapphire laser radiation, the photon energy of the laser was varied in the range from 1.634 to 1.759 eV; the pumping power density was 0.5 W/cm$^2$. The laser radiation was polarized either circularly or linearly, depending on the experiment. The sample was placed in a closed-cycle helium cryostat at a temperature of 12 K.

Using an electromagnet, a magnetic field of up to 400 mT was applied to the sample in Faraday geometry; the direction of the field coincided with the wave vector of the exciting light ($B \parallel k$). PL was detected in the "reflection" geometry, the degree of PL circular or linear polarization was measured, the optical signal was recorded by a spectrometer (spectral resolution was 0.45 nm per 1 mm of slits opening).

The degree of circular polarization $\rho_c$ was determined according to the expression

$$\rho_c = (I^+ - I^-)/(I^+ + I^-), \tag{1}$$

where $I^+$ ($I^-$) is the intensity of the PL component which circular polarization coincides with (is opposite to) the polarization of the exciting light. The degree of linear polarization $\rho_L$ was determined by the expression

$$\rho_L = (I^\parallel - I^\perp)/(I^\parallel + I^\perp), \tag{2}$$

where $I^\parallel$ ($I^\perp$) is the intensity of the PL component linearly polarized along the direction [110] ($[1\bar{1}0]$), when excited by light linearly polarized along [110]. The choice of the crystallographic direction is due to the fact that in the studied (In,Al)As/AlAs QDs the largest degree of exciton alignment is observed along the [110] direction.

## 3. PHOTOLUMINESCENCE SPECTRA. CHARACTERIZATION OF THE SAMPLE

In Ref. [9] it was shown that the ensemble of (In,Al)As/AlAs QDs consists of quantum dots direct band gap and indirect band gap in momentum space, the type of QD is determined by its size (see Fig. 1). As the size of the QDs decreases, the energy of the ground size-quantization level of the electron in the Γ state increases faster than in the X state, since the effective mass of the electron at the Γ point is smaller. Accordingly, when the size of the QD is small enough, the conduction band ground state shifts from the Γ valley to X valley.

PL was excited resonantly with laser energy tuned within the inhomogeneously broadened emission band of the QDs. In this work, we studied an ensemble of indirect band gap (In,Al)As/AlAs QDs located slightly to the right of the Γ-X crossing point in Fig. 1. When a photon is absorbed in such QDs, a hole is excited in the Γ valley, an electron is also predominantly exited in the Γ valley (see blue upward arrow in Fig. 1) due to the difference in the oscillator strength of a direct and indirect exciton. After this, the electron thermalizes to the ground state, located in the X valley, and then recombines with the hole, see the dashed purple downward arrow in Fig. 1.

Fig. 2 shows the PL spectra for seven different energies of excitation photons (further we will use the expression - excitation energy). The PL band corresponding to indirect band gap QDs is shifted from the excitation energy by the $\Delta E_{\Gamma X}$ energy that corresponds to the difference in energy between the Γ and X states (see Fig. 1). Thus, during resonant excitation of indirect gap QDs, the Stokes shift of a different value for different QD sizes is observed. In the spectra, two PL bands of indirect excitons $I_{low}$ and $I_{high}$ can be distinguished (see Fig. 2), which is described in more detail in Ref. [9]. Two indirect bands must correspond to two Γ-X crossing points $E_{\Gamma X}^{low}$ and $E_{\Gamma X}^{low}$ according to Fig. 1. The presence of two Γ-X crossing energies can be caused, for example, by the self-organization of two subensembles of QDs of different compositions. This work examines the $I_{low}$ band, in which the effect of Γ-X mixing on the exciton fine structure is observed.

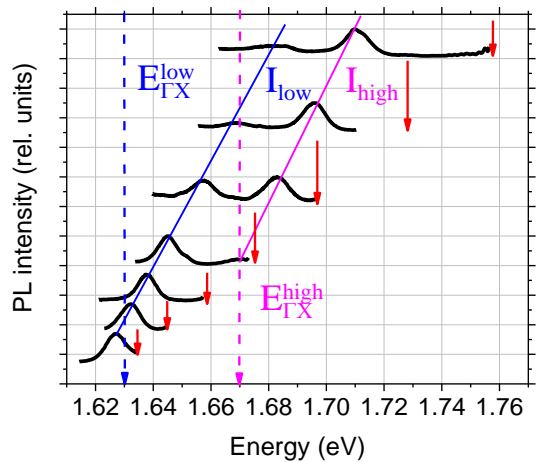

Fig.2. Photoluminescence spectra of an ensemble of indirect band gap (In,Al)As/AlAs QDs depending on the excitation energy. Red arrows indicate laser radiation energies. The experiment was carried out at a temperature of T = 12 K.

4. **EXPERIMENT**

The effect of direct and indirect exciton states mixing on the fine structure was investigated by studying polarized PL in a magnetic field at different excitation energies. Fig. 3(a)/(b) shows the linear/circular polarization of PL subject to magnetic field in Faraday geometry for the case of excitation by linearly/circularly polarized light with a photon energy of 1.66 eV. In magnetic field, the optical alignment of excitons is suppressed (from 35 to 6%), and the optical orientation is restored (from 8 to 33%), which can be described taking into account the fine structure of exciton states.

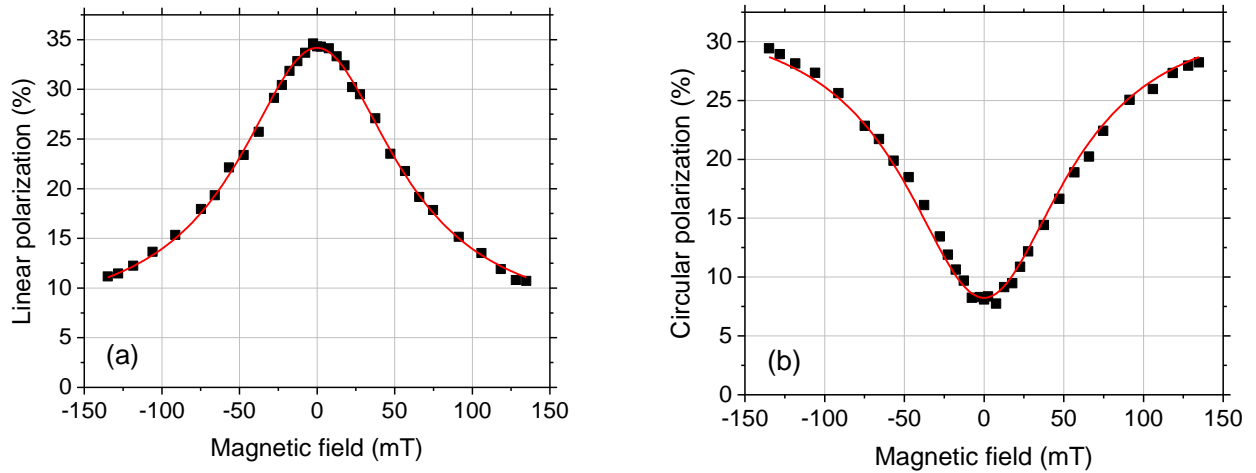

Fig. 3. Dependences of linear (a) and circular (b) PL polarization on the longitudinal magnetic field. Black squares are experimental data, red lines are a description of the data by relations (3) and (4). The energy of excitation is 1.66 eV, detection energy is 1.64 eV, the experiment was carried out at a temperature of T = 12 K.

In direct band gap QDs, the isotropic exchange interaction of an electron and a hole splits the fourfold degenerate states of an exciton consisting of an electron and a heavy hole into bright (optically active) states with projections of the total angular momentum onto the quantization axis $|\pm 1\rangle$ and dark (optically inactive) states $|\pm 2\rangle$ [12]. In the case of deformation in the QD plane, an anisotropic component of the exchange interaction arises, which splits the states of bright excitons into energy $\delta_b$. The $|\pm 1\rangle$ states are split into two states which are linearly polarized in two orthogonal directions x and y, $|X\rangle = \frac{1}{\sqrt{2}}(|+1\rangle + |-1\rangle)$ and $|Y\rangle = \frac{1}{i\sqrt{2}}(|+1\rangle - |-1\rangle)$ [12,13]. In zero magnetic field, linearly polarized light excites pure linearly polarized states $|X\rangle$ or $|Y\rangle$ and the effect of exciton alignment is observed. Circularly polarized light excites a superposition of linear states and the optical orientation signal is observed to the extent of the ratio between the coherence time ($\tau_s = \frac{\hbar}{\delta_b}$) and the exciton lifetime. The above fine structure of levels also allows us to describe the case of an indirect exciton with admixture of direct states, which is described in more detail in the Discussion section.

The magnetic field in Faraday geometry causes the Zeeman splitting, and, when the field is strong enough ($\mu_B g_{ex} B \geq \delta_b$, where $\mu_B$ is the Bohr magneton, $g_{ex}$ is the exciton g factor), linearly polarized states $|X\rangle$ and $|Y\rangle$ are transformed into circular states $|+1\rangle$ and $|-1\rangle$. Linearly polarized light now excites not linearly polarized states, but a superposition of circular states, which may suppress the effect of exciton alignment (Fig. 3(a)). The alignment is suppressed when $\tau \geq \frac{\hbar}{\mu_B g_{ex} B}$. However, the linear polarization saturates at the level of 6% in big magnetic fields, what can be caused by the QDs deformation. Circularly polarized light in Faraday magnetic field excites "pure" states, as a result the optical orientation is restored (Fig. 3(b)). PL circular polarization exceeds 30% in big fields, which it is, however, much less than 100% due to spin relaxation during exciton thermalization process [14].

The dependences of the optical alignment and orientation of excitons on magnetic field were measured at different excitation energies. We recall that the change in the excitation energy results in resonant excitation of the QDs subensemble with a different median size and, as a consequence, with different PL spectrum, see Fig. 2. The description of the magnetic field dependences using corresponding theory is given in the Discussion section. This section presents fitting of the dependences by Lorentz contours: the dependence on the excitation energy of the curves half-widths at half-maximum (HWHM) and of the polarization values in a zero magnetic field were found.

Fig. 4(a) shows the dependence of the degree of optical orientation and alignment of excitons in a zero magnetic field on the excitation energy. With energy increasing, as a consequence increasing shift from the Γ-X mixing point, the circular polarization raises from 3 to 18%, while the linear polarization declines from 53 to 6%. Fig. 4(b) shows the dependence of the HWHM of the magnetic field dependences of polarization on the excitation energy; the half-width was obtained from the fitting of the effects by the Lorentz contour. With energy increasing, the half-widths of restoration of the optical orientation and suppression of exciton alignment decrease from 225 to 20 mT. It should be noted, that at a fixed excitation energy HWHM of the optical orientation restoration and exciton alignment suppression are close to each other. The excitation energy dependences are due to different extent of mixing of direct exciton states with indirect ones, which is discussed in more detail in the following section.

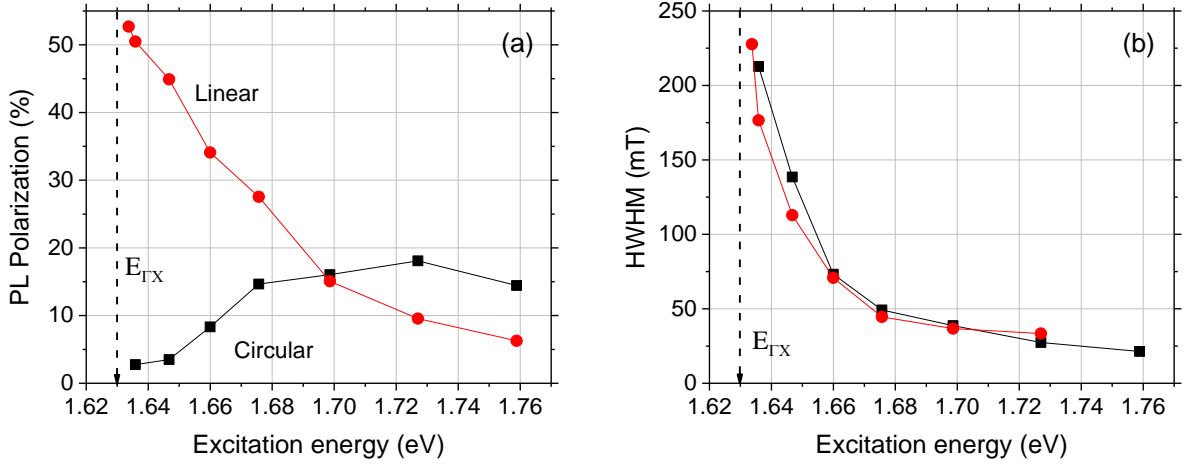

Fig. 4. (a) Dependence of PL polarization on the excitation energy in zero magnetic field. Black squares correspond to the circular polarization of the PL, red circles correspond to linear polarization. (b) Excitation energy dependences of the half-widths of magnetic field polarization dependences at half-maximum. Black squares correspond to optical orientation, and red circles correspond to exciton alignment. PL was detected at the maximum of the corresponding PL spectral bands, see Fig. 2. The energy of the Γ-X crossing point is indicated in the figure by a dashed arrow. The experiment was carried out at a temperature of T = 12 K.

## 5. DISCUSSION

The magnetic field dependences of the PL polarization (see Fig. 3 and Fig. 4) for different excitation energies were described by the expressions for the optical orientation and alignment of a neutral exciton in a magnetic field in Faraday geometry [15,16]:

$$\rho_c(B_\parallel) = \rho_c^0 \frac{1+\Omega_\parallel^2 \tau^2}{1+(\Omega_\parallel^2 + \frac{\delta_b^2}{\hbar^2})\tau^2}, \qquad (3)$$

$$\rho_L(B_\parallel) = \rho_L^0 \frac{1+\frac{\delta_b^2}{\hbar^2}\tau^2}{1+(\Omega_\parallel^2 + \frac{\delta_b^2}{\hbar^2})\tau^2}, \qquad (4)$$

where $\Omega_\parallel = \frac{\mu_B g_x^\parallel B_\parallel}{\hbar}$, the g factor of an optically active exciton is obtained from g factors of an electron and a heavy hole in the X valley, $g_{ex} = g_{hh} - g_e$, $g_{hh} = 2.43$, $g_e = 2$, $g_{ex} = 0.43$ [10]. The parameter $\tau$ denotes the exciton lifetime, $\rho_c^0$ is the maximum degree of circular polarization of the PL in high magnetic fields, $\rho_L^0$ is the degree of linear polarization in zero field. Relations (3) and (4) are suitable for the case of a finite exciton lifetime and the absence of spin relaxation, apart from relaxation that is associated with the electron-hole anisotropic exchange interaction.

It is possible to select the parameters for which the family of experimental curves is accurately described; see the comparison of theory and experiment for the PL excitation energy of 1.66 eV in Fig. 3. The lifetime and anisotropic exchange splitting can be independently determined solely from the description of the optical orientation in a magnetic field by the relation (3). After finding the required parameters for each excitation energy, we checked that substituting the parameters into relation (4) gives an accurate description of the exciton alignment subject to magnetic field. Fig. 5(a) shows the dependence of the desired parameters $\delta_b$ and $\tau$ on the excitation energy.

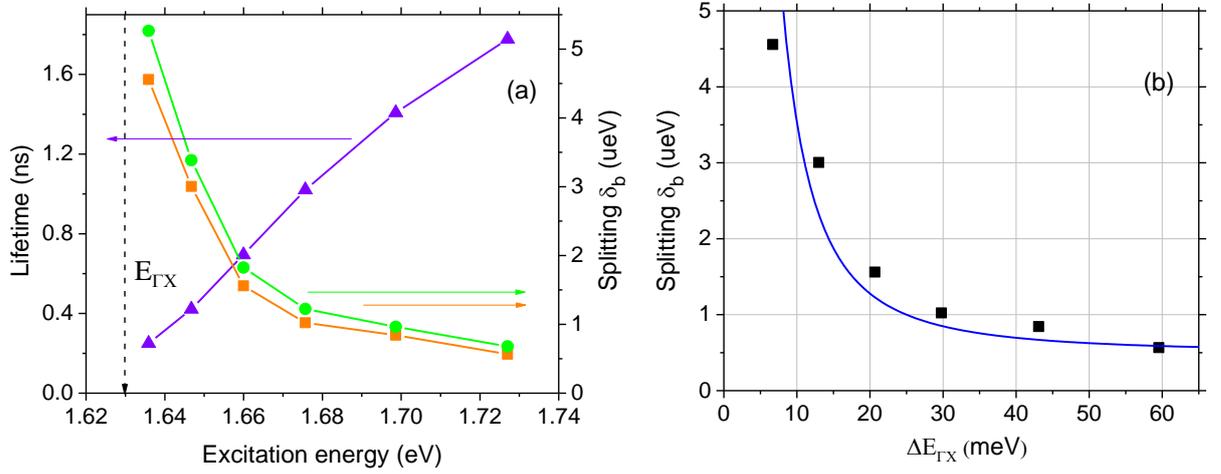

Fig. 5. (a) The dependences of exciton lifetimes (purple triangles) and the magnitude of anisotropic exchange splitting (orange squares and green circles) on the PL excitation energy. The parameters are obtained from the description of experimental data by relation (3). The $\delta_b$ values shown by orange squares correspond to the $\tau$ values shown by purple triangles; the $\delta_b$ values shown by green circles are obtained in the $\tau = \infty$ limit. (b) The dependence of $\delta_b$ on the splitting $\Delta E_{\Gamma X}$ of $|\Gamma\rangle$ and $|X\rangle$ states. The experimental data is shown by black symbols, the fitting using equation (6) is shown by the line.

In Fig. 5(a), one can see an increase in the lifetime with excitation energy (when the energy distance between $|\Gamma\rangle$ and $|X\rangle$ states $\Delta E_{\Gamma X}$ increases). This feature is consistent with what was observed in Ref. [10], where the PL dynamics of an ensemble of (In,Al)As/AlAs QDs was studied. For direct band gap QDs, an exciton lifetime within 1 ns was obtained, for indirect band gap QDs $\tau$ up to 10 μs was observed. Accordingly, in the area of the Γ-X mixing, a transition region

between these values was observed. Thus, the greater the admixture of direct states to indirect excitons, the shorter the lifetime. We note that the $\tau$ (see Fig. 5(a)) is in the range from 0.25 to 1.75 ns, which in order of magnitude corresponds to the times of direct QDs. For "pure" indirect QDs, the $\tau$ in this structure is 200 ns [17]. For indirect excitons with admixed direct states, one would expect values intermediate between 1 and 200 ns. This discrepancy between the expectations and the fitting results is explained by the following. The determination of $\tau$ from the description of the experiment using relation (3) is critically affected by the presence of a constant contribution to the circular polarization of the PL, independent of magnetic field in Faraday geometry. This kind of contribution could be observed, when a part of the QDs in an ensemble are charged with resident electrons [18]. In this case, strictly speaking, the lifetime can take uncertain values in the range from those shown in Fig. 5(a) to (nominally) infinite, so it is not possible to unambiguously determine the absolute values of $\tau$ in the current case. However, it is significant that the arbitrariness in determining the lifetime has virtually no effect on the determination of the anisotropic exchange splitting. Fig. 5(a) shows that $\delta_b$ obtained for infinite lifetimes (green circles) differ by no more than 20% from the values $\delta_b$ (orange squares) obtained with the smallest realizable lifetimes. Thus, $\delta_b$ is reliably determined, the values can be found in the range between the green and red symbols in Fig. 5(a).

At a high excitation energy (1.727 eV), the small splitting of the optically active exciton levels is observed ($\delta_b = 0.57..0.68$ µeV), which is on the order of the splitting in "purely" indirect QDs ($\delta_b \leq 0.2$ µeV). As the PL excitation energy decreases, the admixture of direct states increases and, as a consequence, $\delta_b$ increases; when approaching the Γ-X crossing point ($E_{exc} =$ 1.636 eV), the splitting reaches a value determined in the range $\delta_b \approx 4.56..5.27$ µeV. At the same time, in "pure" direct QDs (In,Al)As/AlAs $\delta_b$ is on the order of 100 µeV [8,11]. Note that the region of Γ-X mixing occupies an energy range of the order of tens of meV around the Γ-X crossing point (1.63 eV).

Here, we present a simple model of Γ-X mixing. The wave function of exciton in the case of the system with two orthonormalized basis states $|\Gamma\rangle$ and $|X\rangle$ is given by

$$\Psi = C_\Gamma |\Gamma\rangle + C_X |X\rangle, \qquad (5)$$

The bright exciton anisotropic exchange splitting of the mixed state can be expressed as

$$\delta_b = \delta_b^\Gamma |C_\Gamma|^2 + \delta_b^X |C_X|^2, \qquad (6)$$

Where $\delta_b^\Gamma$ ($\delta_b^X$) is the exchange splitting of "pure" direct(indirect) exciton, the coefficients are given by

$$|C_\Gamma|^2 = \frac{1}{2}(1 - \frac{\Delta E_{\Gamma X}}{\sqrt{\Delta E_{\Gamma X}^2 + 4|V_{\Gamma X}|^2}}), \quad (7)$$

$$|C_X|^2 = \frac{1}{2}(1 + \frac{\Delta E_{\Gamma X}}{\sqrt{\Delta E_{\Gamma X}^2 + 4|V_{\Gamma X}|^2}}), \quad (8)$$

where $|V_{\Gamma X}|$ is the modulus of the matrix element of the coupling between the $|\Gamma\rangle$ and $|X\rangle$ states. The experimental data of $\delta_b$ depending on the splitting $\Delta E_{\Gamma X}$ of $|\Gamma\rangle$ and $|X\rangle$ states was fit using the equation (6), see Fig. 5(b). The data is in good agreement with the theory, the following parameters were used: $\delta_b^\Gamma = 260$ μeV [8], $\delta_b^X = 0.5$ μeV, $V_{\Gamma X} = 1.1$ meV. The last parameter allows to estimate the Γ-X coupling strength.

Let us consider now the PL polarization dependence on the excitation energy. From Fig. 4(a) it can be seen that at high energies the degree of exciton alignment in zero magnetic field is small (6%) and a relatively large degree of optical orientation is observed (18%), which can be explained in terms of the exciton fine structure in indirect QDs. The exchange splitting $\delta_b$ in such excitons is small, as a result the optically oriented spin orientation is conserved during the exciton lifetime. As the excitation energy decreases, the $\delta_b$ increases, and the optical properties of such exciton increasingly correspond to those characteristic for direct QDs. In particular, near the Γ-X crossing point at a PL excitation energy of 1.636 eV, the admixture of direct states is high and the coherence of the states $|X\rangle$ and $|Y\rangle$ excited by a circularly polarized photon is lost in time $\frac{\hbar}{\delta_b} \approx 0.13$ ns $< \tau$. Thus, a low degree of PL optical orientation is observed (3%). The linearly polarized photon excites the "pure" state $|X\rangle$ or $|Y\rangle$, resulting in the large degree of linear polarization reaching 53% at an energy of 1.636 eV.

## 6. CONCLUSION

The properties of the ensemble of (In,Al)As/AlAs QDs were studied subject to the QD size using the method of polarized PL in a Faraday magnetic field. It was shown that the admixture of direct states changes the fine structure of indirect in k-space excitons, significantly increasing the anisotropic exchange splitting of optically active excitonic levels. Admixing direct states to the indirect ones allows one to smoothly change the optical properties of excitons – to a great extent suppressing the optical orientation and largely restoring the alignment of excitons.

## ACKNOWLEDGMENTS

The authors are grateful to D. S. Smirnov for fruitful discussions. We acknowledge the support by the Russian Science Foundation (Grant No. 22-12-00125).